\documentstyle[preprint,prl,aps]{revtex}
\begin{document}
\preprint{\today}
\draft
%
%
\title{Detecting Casimir Forces\\
through a Tunneling Electromechanical Transducer}
\author{Roberto Onofrio${}^{1,2}$ and Giovanni Carugno${}^{2}$}
\address{${}^{1}$Dipartimento di Fisica ``G. Galilei'',
Universit\`a di Padova, Via Marzolo 8, Padova, Italy 35131\\
${}^{2}$Istituto Nazionale di Fisica Nucleare, Sezione di Padova, Italy 35131
}
\date{\today}
\maketitle
%
%
\begin{abstract}
We propose the use of a tunneling electromechanical transducer to dinamically
detect Casimir forces between two conducting surfaces.
The maximum distance for which Casimir forces should be detectable
with our method is around $1 \mu$m, while the lower limit is given by the
ability to approach the surfaces.
This technique should permit to study gravitational
forces on the same range of distances, as well as the vacuum friction
provided that very low dissipation mechanical resonators are used.
\end{abstract}
%
%
\pacs{03.70.+k, 06.70.Mx, 73.40.Gk}
%
%
\narrowtext
One of the most astonishing and less understood consequences of the existence
of quantum vacuum is the possibility to observe forces between macroscopic
neutral bodies, as outlined in a pioneering paper by Casimir \cite{CAS1}.
Casimir forces are of purely relativistic origin and this is manifested
by their retarded nature, thus vanishing at very small distances.
The force per unit area between two conducting, but neutral,
indefinite and  parallel plates has been calculated as
\begin{equation}
P_C={K_C\over d^4},
\label{CAS1}
\end{equation}
where $d$ is the distance between the two plates, assumed greater
than the penetration lenght in the conductors $\simeq 100$nm, and
$K_C={\pi h c/480}=1.3 \cdot 10^{-27} N m^2$
is a purely kinematic constant depending
upon the speed of light $c$ and the Planck constant $h$.
Until now few experiments
have been performed to verify this simple law.
The submicron region has been explored in the range $0.5\div 2 \mu$m
\cite{SPAR0,SPAR1,SARL}.
Measurements have also been reported
in the range $50\div 130$nm  \cite{ISRA} while a proposal
aimed at observing Casimir forces in the cm range using optical techniques
has been recently discussed in \cite{JACO}.
In this Letter we propose a detection technique
for macroscopic forces and we report on preliminary
results concerning the calibration of a prototype for the observation
of Casimir forces by means of a tunneling electromechanical transducer.
Our estimates suggest that the Casimir force should be detectable
at least in a range of two decades below $10 \mu$m.

The detection scheme is schematically shown in Fig. 1.
The two parallel conducting surfaces are made of
a cantilever beam with a mass of circular shape at its extreme, rigidly
clamped to the other end on a base in such a way to oscillate,
and a disk rigidly connected to a piezoelectric stack.
The distance between these two surfaces can be changed both
by driving the mechanical oscillator at its proper frequency $\nu_r$
through a piezoelectric gauge and by driving the disk
at a fixed frequency $\nu_p$ through the piezoelectric stack.
For practical reasons the vibration frequency
of the piezoelectric stack is smaller
than the one of the mechanical resonator.
On the other side of the resonator a tunneling electromechanical
transducer allows to measure the amplitude of its oscillations.
When the disk approaches the resonator the attractive
Casimir force should appear and
a static displacement should result in a decrease of the tunneling current
detected on the other side of the resonator.
However this measurement is affected by all the usual drawbacks
of a static detection and this is true also for a very low-frequency
detection scheme due the presence of $1/f$-like
noise of mechanical and electrical nature.
The heterodyne detection, as discussed in detail in \cite{HUNK}
is unaffected by this problem: let us suppose that both
the resonator and the disk are sinusoidally driven
at the angular frequencies
$\omega_r=2\pi \nu_r$ and $\omega_p=2 \pi \nu_p$
and the amplitudes of the corresponding
oscillations are respectively $x_r$ and $x_p$.
Thus the gap between the two plates will become time-dependent as
\begin{equation}
d(t)=d_0+x_r\cos(\omega_rt)+x_p\cos(\omega_pt).
\label{GAPT}
\end{equation}
The Casimir force will be therefore modulated and, by supposing that
the infinite plane is a valid approximation for a plate of finite
surfaces, its expression will be written as
\begin{equation}
F_C(t)\simeq {K_C S\over{d_0^4[1+(x_r/d_0)\cos\omega_rt+
(x_p/d_0)\cos\omega_pt]^4}}.
\label{GAPT1}
\end{equation}
If we suppose a weak modulation regime in (\ref{GAPT}), i.e. $x_r,x_p<<d_0$,
we obtain, up to the third order in $x_r/d_0,x_p/d_0$
\begin{equation}
F_C(t)=K_C{S\over d_0^4} [1+g(t)],
\label{GAPT2}
\end{equation}
where $g(t)$ has Fourier components shown in Tab. I.
These Fourier components of the Casimir force give rise to Fourier
components for the motion of the mechanical resonator at the same
frequencies, which are finally detected by the tunneling tip.
The relationship between force and displacement
is given by the mechanical transfer function of the resonator.
When the disk is put apart enough from the resonator only the component
driven at mechanical resonance $\omega_r$ survives.
The other linear component at $\omega_p$ can be also exploited to
get information on the presence of forces but an accurate shielding from
the vibrations induced in the resonator by the PZT stack is required.
Moreover the presence of 1/f noise
in the tunneling current limits the possibility of exploiting this
low-frequency harmonic as well as the other one at $2\omega_p$.
The existence of a very small modulation parameter $x_r$,
being for practical reasons $x_r<<x_p$, provides rationale
for ruling out the Fourier component at $2\omega_r$, which should
be also affected by the non-linear relationship between tunneling current
and tip-sample gap.
The lateral bands around the resonant frequency
seem the best candidates to obtain information on the Casimir force, being
less plagued by seismic noise or electric 1/f noise
in the tunneling transducer.
On the other hand these two components are suppressed
by a factor $10 x_r x_p/d_0^2$ with respect to the static case.

The considerations made above are quite conservative
and assume for instance that no accurate acoustic insulation has been
provided to prevent acoustic pick-up at $\omega_p$ due to the PZT stack.
Less pessimistic considerations can only enforce our
estimates also by using a redundancy among the informations
present at the different Fourier components.

This detection scheme has been tested through a
first calibration by biasing
at constant voltage the gap formed between the resonator and the disk.
The apparatus used is schematically shown in Fig.2.
The tunneling electromechanical transducer is similar
to the one already described in \cite{CARU}
and its sensitivity is of the order of $10^{-11}$m at a frequency of $60$KHz.
Its main advantages with respect to other types of transducers are the
possibility of monitoring small masses with enough sensitivity and the
absence of problems related to the parallelism, the probe being point-like.
Unlike in ref. \cite{CARU} the approach of the tip to the surface
of the resonator is obtained through an inchworm motor \cite{INCH}.
The resonator is made out of a single piece of
steel machined through electroerosion
and its vibrating mass is a cylinder with diameter of $6$mm and
a thickness of $0.2$mm. The  lenght of
the cantilever beam, with a rectangular section of $0.1$mm times
$0.2$mm, is equal to $4$mm. A schematic view of the resonator is
shown in the upper-left part of Fig.3.
In the same figure the mechanical transfer
function, obtained by the FFT  of the tunneling
current under white noise mechanical excitation, is shown.
A frequency for the first flexural mode of the cylinder
$\nu_r=19$KHz and a mechanical quality factor of $10^2$ are measured.
The piezoelectric stack has a distance modulation per unit
of voltage equal to $0.2\mu$m/V with a high-frequency cut-off around $5$KHz.
It is connected to a flat surface of Aluminum, 10 mm diameter, 4 mm thick.
The apparatus is put on a four cantilever beam suspension similar
to the ones used for suspension of gravitational wave detectors \cite{COCC}.
The disk is biased at constant voltage V and the resulting
force acting on the resonator is, up to second order,
\begin{equation}
F_V(t)={{CV^2}\over 2d(t)}\simeq {\epsilon_0 S\over 2} \bigl({V\over d_0}
\bigr)^2
(1-{x_r\over d_0}\cos \omega_r t-{x_p\over d_0}\cos \omega_p t)^2,
\label{GAPT3}
\end{equation}
which contains also terms at the lateral bands $\omega_r\pm\omega_p$
of amplitude equal to
\begin{equation}
F_V(\omega_r\pm \omega_p)={\epsilon_0 S \over 2} \bigl({V\over d_0}\bigr)^2
{x_r \over d_0} {x_p \over d_0}.
\label{GAPT4}
\end{equation}
This has to be compared with the component of the Casimir
force at the lateral bands as deduced from Eqn.4 and Table I:
\begin{equation}
F_C(\omega_r\pm \omega_p)=10 K_C {S \over d_0^4} {x_p\over d_0}
{x_r \over d_0}.
\label{GAPT5}
\end{equation}
To get an estimate of the sensitivity it is useful to
introduce an equivalent voltage $V_{eq}$ such that
for a given distance $d_0$
a bias equal to $V_{eq}$ simulates the Casimir force.
This equivalent voltage is independent
of the modulation parameters and, by equating  Eqns. (\ref{GAPT4}) and
(\ref{GAPT5}), we get
\begin{equation}
V_{eq}={\bigl({20K_C\over \epsilon_0}\bigr)}^{1/2}
{1\over d_0}={5.42\cdot 10^{-8}\over
d_0},
\label{VEQ}
\end{equation}
which means, for instance, that the Casimir force at distance $d_0=10^{-6}$m
can be simulated by a bias voltage $V_{eq}\simeq 50$mV.
Lateral bands due to the electric field are observed in Fig. 4b,c which show
their appearance only when the bias voltage and the driving
voltage for the piezoelectric stack are both present.
By comparison we also report in Fig.4a the spectrum
observed in absence of the bias voltage.
The two cases of Fig. 4b,c are relative to $V=150$V and $V=300$V
respectively.  In the latter case the lateral peaks
are above the noise by a factor 6.
{}From this we deduce that a signal to noise ratio SNR=1
corresponds to $V_{eq}^{SNR=1}\simeq 70$V.
This has to be compared with
the equivalent voltage of the Casimir force at $d_0=10^{-6}$m
which is three orders of magnitude smaller.
Many improvements can be made on a dedicated prototype to increase
the sensitivity by three orders of magnitude.
Resonators with larger mechanical
quality factors and operating in vacuum can be used.
The wide-band noise can be reduced by optimizing the electronic chain.
Finally, phase sensitive detection techniques can be exploited.
The first tests show that more than a factor 10 in the sensitivity
is gained by using the vector averaging mode of the FFT spectrum analyzer.
The ultimate limit on the reduction of the noise uncorrelated
with the signal is determined in this last case by the
stability of the tunneling current during the integration time, which
we observe to be constant at least on a scale of tens of minutes.

The lower limit of the distance at which Casimir forces are measurable is
given by the ability to control the distance between the two plates.
It is possible to put the previous apparatus
inside an electron microscope and to use piezoelectric
quadrant tubes to control the parallelism.
In this way uniform gaps of the order of 100nm
or less can be obtained, allowing for a quantitative
test within two orders of magnitude of the distance.
Flat surfaces on the scale of less than 100nm can be obtained
exploiting polishing techniques already used in optical devices.
Preliminary tests have been performed on microresonators
and micrometers to check the feasibility of a monitoring of surfaces gaps
in the $100$nm range inside an electron microscope.

Besides the study of the Casimir force we believe this
detection technique could also be exploited for the measurement
of other forces spatially dependent on the same range of distances,
such as gravitational forces.
In this case particular configurations should be used for
the geometry of the test masses to allow very small
distances between the center of mass of the bodies.
The dependence upon the distance can be also determined by
an accurate measurement of the Fourier components at a given distance.
By repeating the previous considerations for a generic force
of the type
\begin{equation}
F_n={K \over d^n}
\label{GAPTN}
\end{equation}
we get a dynamical force
\begin{equation}
F_n(t)={K\over d_0^n} [1+g_n(t)],
\label{GAPTN1}
\end{equation}
where $g_n(t)$ has Fourier components, up to the second order, shown in
Tab. II.
\noindent
The various components depend upon n, which in turn can be measured
by determining for instance the ratio
\begin{equation}
G_n(\omega_p)/G_n(\omega_r\pm \omega_p)={2\over n+1}{d_0\over x_r}
\end{equation}
where $d_0/x_r$ can be determined using a simultaneous
calibration with an electric field.
To observe the Fourier component at $\omega_p$ an accurate shielding
from the vibration induced by the PZT stack is mandatory.
For n of physical interest (gravitational
forces, Casimir forces, i.e. n=2,4)
the differences in the values are above 20 per cent and should be detectable,
opening also the possibility to study gravity on $\mu$m distances with
geometries of the bodies such that the center of mass can be
spaced by such order of magnitude.
The static gravitational force between the masses used in our prototype is
$F_G\simeq 6\cdot 10^{-7}$ N
supposing that the centers of mass are spaced by $1\mu$m, {\it i.e.}
one order of magnitude larger that the corresponding Casimir force.

The proposed detection method is also valid far from the mechanical resonance.
However the use of a resonance condition is welcome
because of the gain, equal to the mechanical
quality factor, with respect to a non-resonant configuration.
Use of microresonators with very high mechanical quality factors
could also allow for an alternative detection scheme based upon the
measurement of the resonant frequency shift induced by the Casimir force.
Furthermore, a resonance condition has been recently
suggested \cite{BRAG2}, \cite{BRAG3} to detect the dissipative
part of the Casimir force as predicted in \cite{LEVI}.
Although this requires very high mechanical quality
factors, the improvements made in the technology of
microresonators \cite{KAMI}, \cite{BUSE}
allow to hope that in a close future the dissipative
nature of the quantum vacuum could be detected and studied.

\acknowledgments
We acknowledge V. B. Braginsky for fruitful discussions and
A. Bettini, M. Cerdonio, E. Conti, F. Illuminati and
G. Ruoso for a critical reading of the manuscript.
%
%

%
%
\begin{table}
\begin{center}
\begin{tabular}[t]{|c|c|}
$\omega$ & $G(\omega)$\\
\hline
$\omega_p$             & $4 x_p/d_0$       \\
$2\omega_p$            & $5 (x_p/d_0)^2$   \\
$\omega_r- \omega_p$   & $10 x_p x_r/d_0^2$\\
$\omega_r$             & $4 x_r/d_0$       \\
$\omega_r+ \omega_p$   & $10 x_p x_r/d_0^2$\\
$2\omega_r$            & $5(x_r/d_0)^2$    \\
\end{tabular}
\caption{Fourier components of the dynamical term $g(t)$}
\end{center}

\begin{center}
\begin{tabular}[t]{|c|c|}
$\omega$ & $G_n(\omega)$\\
\hline
$\omega_p$             & $n x_p/d_0$                     \\
$2\omega_p$            & $\frac{n(n+1)}{4} (x_p/d_0)^2$  \\
$\omega_r- \omega_p$   & $\frac{n(n+1)}{2} x_p x_r/d_0^2$\\
$\omega_r$             & $n x_r/d_0$                     \\
$\omega_r+ \omega_p$   & $\frac{n(n+1)}{2} x_p x_r/d_0^2$\\
$2\omega_r$            & $\frac{n(n+1)}{4}(x_r/d_0)^2$   \\
\end{tabular}
\caption{Fourier components of the dynamical term $g_n(t)$}
\end{center}
\end{table}
%
\begin{figure}
\caption{Dynamical detection scheme for small forces
acting on a micromechanical resonator.}
\end{figure}

\begin{figure}
\caption{Scheme of the apparatus used in the calibration.}
\end{figure}

\begin{figure}
\caption{FFT spectrum of the tunneling current under white-noise
excitation of the resonator whose shape is drawn in the upper-left corner.
The curve is proportional to the transfer function of the
mechanical oscillator, and allows to measure
resonant frequency and mechanical quality factor.}
\end{figure}

\begin{figure}
\caption{FFT Spectrum of the tunneling current
in presence of a mechanical modulation of the resonator
without (a) and with (b,c) a mechanical modulation of the disk under biasing
at constant voltage respectively at $V=150$V and $V=300$V.
Lateral bands at $\omega_r\pm\omega_p$ are observable with
amplitude depending both upon the driving voltage of the piezoelectric
stack and the bias voltage.
The lateral bands in b) are about 12 dB below the
ones in c) which corresponds to a quadratic scaling for the signal
versus the bias voltage. In the case  c) we have
$\nu_r=19.55$KHz, $\nu_p=1$KHz, $V=300$V, $d_0=10^{-4}$m,
$x_r \simeq 10^{-9}$m, $x_p=2\cdot 10^{-7}$m, $S=2.83 \cdot 10^{-5}m^2$
which implies a Fourier component of the force
$F_V(\omega_r\pm \omega_p)\simeq 2\cdot 10^{-11}$N.}
\end{figure}

\end{document}